\documentclass[aps, pra, superscriptaddress, nofootinbib,
               amsmath, amssymb, twocolumn,
               preprintnumbers, showpacs,
               raggedbottom,
               floatfix,
               todo,
              ]{revtex4-1}
\usepackage[utf8]{inputenc}
\usepackage[aps]{my_paper}                      
\usepackage{my_acronyms}                         
\usepackage{float}

\exclude{
A B C D E F G H
Related Videos:
* http://youtu.be/VxvOPsOs-6s: Medium anisotropy, no anharmonicity
* http://youtu.be/uDuOOVOO5D4:   ETF analogue
* http://youtu.be/gIfhB7QrGxU: Small anisotropy, no anharmonicity
* http://youtu.be/VUrlvJglqkI:   ETF analogue
* http://youtu.be/7qG03kmN-fs: Medium anisotropy and anharmonicity
* http://youtu.be/L5ONtMQDfgE:   ETF analogue
* http://youtu.be/K74GUp1cDnM: Small anisotropy and anharmonicity
* http://youtu.be/HOIB-R1TwJU:   ETF analogue
* http://youtu.be/0pg8N91-gZ0: Medium (anti)anisotropy and anharmonicity
* http://youtu.be/EXLOlBbsKS8:   ETF analogue
* http://youtu.be/_aWO0tIOOdw: Tilted imprint (no asymmetry)
* http://youtu.be/utpkynd1Qkc: Two tilted imprints (no asymmetry)
* http://youtu.be/7xlG15TUNQs: Rotated imprint with medium anisotropy
* http://youtu.be/nCGtq8k6SAQ: Quantum turbulence
}

\newcommand\movieS{qtxD1VCk-08}
\newcommand\movieA{VxvOPsOs-6s}
\newcommand\movieB{gIfhB7QrGxU}
\newcommand\movieC{7qG03kmN-fs}
\newcommand\movieD{K74GUp1cDnM}
\newcommand\movieE{0pg8N91-gZ0}
\newcommand\movieF{_aWO0tIOOdw}
\newcommand\movieG{utpkynd1Qkc}
\newcommand\movieH{7xlG15TUNQs}
\newcommand\movieETFA{uDuOOVOO5D4}
\newcommand\movieETFB{VUrlvJglqkI}
\newcommand\movieETFC{L5ONtMQDfgE}
\newcommand\movieETFD{HOIB-R1TwJU}
\newcommand\movieETFE{EXLOlBbsKS8}
\newcommand\movieQT{nCGtq8k6SAQ}
\newcommand\youtube[2][0]{
  \href{http://www.youtube.com/watch?v=#2&t=#1}}
\newcommand\utube[1]{
  \url{http://youtu.be/#1}}
\newcommand\allmovies{%
  \textbf{\Gls{SLDA} Movies}\\
  \utube{\movieS}~(Fig.~\ref{fig:science}),\\
  \utube{\movieA}~(Fig.~\ref{fig:strong_asym}t,~$\delta$:~10\%,~$C$:~0),\\
  \utube{\movieC}~(Fig.~\ref{fig:strong_asym}b,~$\delta$:~10\%,~$C$:~3\%),\\
  \utube{\movieB}~(Fig.~\ref{fig:weak_asym}t,~$\delta$:~0.5\%,~$C$:~0),\\
  \utube{\movieD}~(Fig.~\ref{fig:weak_asym}b,~$\delta$:~0.5\%,~$C$:~0.15\%),\\
  \utube{\movieE}~(Fig.~\ref{fig:alignment}t,~$\delta$:~-10\%,~$C$:~3\%),\\
  \utube{\movieH}~(Fig.~\ref{fig:alignment}b,~$\delta$:~10\%,~$C$:~3\%),\\
  \utube{\movieF}~(tilted imprint),\\
  \utube{\movieG}~(colliding vortices),\\
  \utube{\movieQT}~(Fig.~\ref{fig:qt_gen}, turbulence),\\
  \textbf{\Gls{ETF} Movies} (corresponding to the listed figures)\\
  \utube{\movieETFA}~(Fig.~\ref{fig:strong_asym}t,~$\delta$:~10\%,~$C$:~0),\\
  \utube{\movieETFC}~(Fig.~\ref{fig:strong_asym}b,~$\delta$:~10\%,~$C$:~3\%),\\
  \utube{\movieETFB}~(Fig.~\ref{fig:weak_asym}t,~$\delta$:~0.5\%,~$C$:~0),\\
  \utube{\movieETFD}~(Fig.~\ref{fig:weak_asym}b,~$\delta$:~0.5\%,~$C$:~0.15\%),\\
  \utube{\movieETFE}~(Fig.~\ref{fig:alignment},~$\delta$:~-10\%,~$C$:~3\%)}

\graphicspath{{frames/}}

\begin{document}

\title{Life Cycle of Superfluid Vortices and Quantum Turbulence in the Unitary
  Fermi Gas}

\author{Gabriel Wlaz\l{}owski}
\email{gabrielw@if.pw.edu.pl}
\affiliation{Faculty of Physics, Warsaw University of Technology,
  Ulica Koszykowa 75, 00-662 Warsaw, Poland}
\affiliation{Department of Physics, University of Washington, Seattle,
  Washington 98195--1560, USA}

\author{Aurel Bulgac}
\email{bulgac@uw.edu}
\affiliation{Department of Physics, University of Washington, Seattle,
  Washington 98195--1560, USA}

\author{Michael McNeil Forbes}
\email{mforbes@alum.mit.edu}
\affiliation{Department of Physics \& Astronomy, Washington State University,
  Pullman, Washington 99164--2814, USA}
  \affiliation{Institute for Nuclear Theory, University of Washington,
  Seattle, Washington 98195--1550, USA}
\affiliation{Department of Physics, University of Washington, Seattle,
  Washington 98195--1560, USA}

\author{Kenneth J. Roche}
\email{k8r@u.washington.edu}
\affiliation{Pacific Northwest National Laboratory,
  Richland, Washington 99352, USA}
\affiliation{Department of Physics, University of Washington, Seattle,
  Washington 98195--1560, USA}

\date{\today}

\newcommand{\error}[1]{\marginpar{\color{red}\tiny #1}}
\newcommand{\acite}[1]{%
  \ifcsdef{cite#1}
  {\cite{\csuse{cite#1}}}
  {\cite{???}\error{#1 undefined}%
    \message{LaTeX Warning: cite#1 not defined.}%
    }}

\newcommand{\citesnake}{Anderson:2001,Brand:2002,Berloff:2002,Komineas:2007}
\newcommand{\citetube}{Komineas:2007}
\newcommand{\citeturbulence}{Vinen:2002,Vinen:2006,Vinen:2010,Skrbek:2011,Tsubota:2008,Tsubota:2013,Paoletti:2011}

\begin{abstract}\noindent%
  The \gls{UFG} offers an unique opportunity to study quantum turbulence both
  experimentally and theoretically in a strongly interacting fermionic
  superfluid.  It yields to accurate and controlled experiments, and admits the
  only dynamical microscopic description via time-dependent \gls{DFT} -- apart
  from dilute bosonic gases -- of the crossing and reconnection of superfluid
  vortex lines conjectured by Feynman in 1955 to be at the origin of quantum
  turbulence in superfluids at zero temperature.  We demonstrate how various
  vortex configurations can be generated by using well established experimental
  techniques: laser stirring and phase imprinting. New imagining techniques
  demonstrated by the MIT group [Ku~\textit{et~al.}
  \href{http://arxiv.org/abs/1402.7052}{arXiv:1402.7052}] should be able to
  directly visualize these crossings and reconnections in greater detail than
  performed so far in liquid helium.  We demonstrate the critical role played
  by the geometry of the trap in the formation and dynamics of a vortex in the
  \gls{UFG} and how laser stirring and phase imprint can be used to create
  vortex tangles with clear signatures of the onset of quantum turbulence.
\end{abstract}

\preprint{NT@UW-14-08}
\preprint{INT-PUB-14-009}

\maketitle
\glsresetall
\glsunset{GPU}
\glsunset{MIT}
\setlength\marginparwidth{40pt}
\noindent

\lettrine{Q}{uantized vortices} are a hallmark of superfluids.  Their
generation, dynamics, evolution, and eventual decay have been studied
experimentally and theoretically for some six decades in liquid $^4$He and
$^3$He, Bose and Fermi cold atom systems, neutron stars, condensed matter
systems, cosmology, and particle physics. The two-component \gls{UFG} is of
particular interest due to precise experimental control in cold atom traps, and
almost direct applicability to dilute matter in neutron star crusts where
experiments and direct observations are not possible. The experimental control
and universality of the system make it one of enormous interest for those
studying phenomena in relativistic heavy-ion collision, nuclear physics,
nuclear astrophysics, atomic physics, and condensed matter physics.

In a recent letter~\cite{Bulgac:2013d}, we showed that the initial conditions as
described in the \gls{MIT} experiment~\cite{Yefsah:2013} -- which claimed to
have an axially symmetric trap -- lead to the production of superfluid vortex
rings.  The properties of these rings provided a natural explanation of several
puzzling characteristics displayed by the objects observed in~\cite{Yefsah:2013}
which they called ``heavy solitons''. In particular, the vortex ring scenario
explained successfully the long oscillation periods, the unusually large
apparent effective mass of the objects, and details of the imaging procedure
that resulted in objects appearing much larger than the natural width of a
vortex.

The experiment~\cite{Yefsah:2013} was recently updated~\cite{Ku:2014} with an
improved method of imaging slices from the original experiment (tomography),
conclusively demonstrating that the ``heavy solitons'' are ultimately vortex
segments that consistently align themselves across the vertical imaging
axis. The authors of~\cite{Ku:2014} suggest that this systematic alignment is
due to asymmetries in the trap arising from gravitational distortions of the
optical trapping potential in the vertical direction.

The optical trapping potential in the $x$ and $y$ directions is an axially
symmetric gaussian altered by gravity in the vertical direction $y$:
\begin{multline}
  V(x,y,z) = \frac{m\omega_z^2z^2}{2} + \order(z^4) + \\
  + V_0\left[1-\exp\left(-\frac{m\omega_x^2(x^2+y^2)}{2V_0}\right)
  \right] + mgy.
\end{multline}
Shifting $y\rightarrow y+y_0$ where $y_0$ is the new minimum gives the following
effective trapping potential
\begin{multline}
  V(x,y+y_0,z) \approx
  \frac{m\omega_z^2z^2}{2} +\order(z^4) + \\
    \frac{m\omega_x^2 x^2}{2} + \frac{m\omega_y^2 y^2}{2}
  + Cy^3 + \order(\delta^2) + \text{const} \label{eq:V_gravity}
\end{multline}
where $\delta = 3mg^2/4\omega_x^2 V_0$ is treated perturbatively, and
\begin{equation}
  \omega_y\approx  \omega_x\left(1 - \delta \right), \quad
  C \approx \frac{2m\omega_x^4}{3g}\delta.
\end{equation}
According to~\cite{Ku:2014}, $\delta = 1-\omega_y/\omega_x \approx 5\%$ is
small, justifying this expansion.  The axial symmetry is thus broken by two
effects: an anisotropy $\delta$ and an anharmonicity $Cy^3$ which we
characterize in terms of $C_0 = m\omega_y^2/2R$ where $C=C_0$ would give equal
quadratic and cubic terms at the \gls{TF} radius $R$ where $V(0, R+y_0, 0) =
\mu$.  With the experimental parameters~\cite{Yefsah:2013, Ku:2014}, $C/C_0
\approx \delta \approx 5\%$.

Here we show that accounting for gravity indeed induces the initially produced
vortex rings~\cite{Bulgac:2013d, Reichl:2013, Wen:2013} to convert into vortex
lines, which oscillate along the long axis of the trap.  The conversion of
off-axis vortex rings into vortex lines on the boundary of a trap was first
demonstrated in the \gls{UFG} in the simulations of~\cite{Bulgac:2011b}
(reproduced in Fig.~\ref{fig:science}).  In the context of~\cite{Yefsah:2013},
the possibility of this conversion was also suggested in
Ref.~\cite{Scherpelz:2014}, where asymmetries were induced manually through
stochastic noise.  Without a systematic asymmetry in the trap, however,
stochastic variations lead to randomly orientated vortex lines inconsistent
with the experimental observations. The simulations~\cite{Scherpelz:2014}
model only the condensate, but lack averaging over stochastic trajectories or a
density matrix as required to properly describe thermal fluctuations.  The role
of these fluctuations also diminishes at low temperatures, becoming irrelevant
at zero temperature. Our zero-temperature approximation is supported
by~\cite{Allen:2014} which finds thermal effects to be negligible for vortex
reconnection.

\let\themovie\movieS
\newcommand\movieref[1][0]{\youtube[#1]{\themovie}}
\begin{figure}[tb]
  \href{http://www.youtube.com/watch?v=qtxD1VCk-08&t=78}{\includegraphics{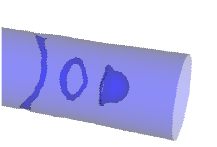}}%
  \href{http://www.youtube.com/watch?v=qtxD1VCk-08&t=86}{\includegraphics{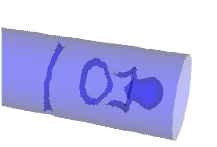}}%
  \href{http://www.youtube.com/watch?v=qtxD1VCk-08&t=92}{\includegraphics{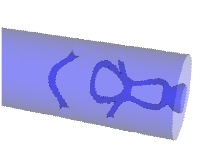}}%
  \href{http://www.youtube.com/watch?v=qtxD1VCk-08&t=98}{\includegraphics{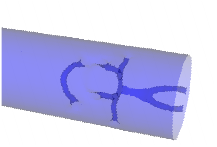}}%
  \\[-1pt]
  \href{http://www.youtube.com/watch?v=qtxD1VCk-08&t=104}{\includegraphics{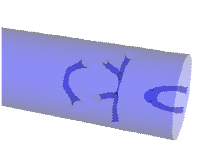}}%
  \href{http://www.youtube.com/watch?v=qtxD1VCk-08&t=110}{\includegraphics{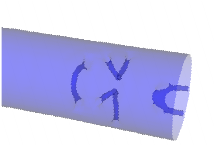}}%
  \href{http://www.youtube.com/watch?v=qtxD1VCk-08&t=116}{\includegraphics{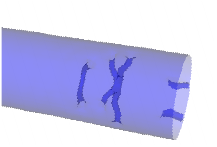}}%
  \href{http://www.youtube.com/watch?v=qtxD1VCk-08&t=122}{\includegraphics{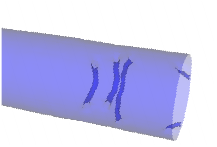}}
  \caption{\label{fig:science}(Color online, click on frames to view movie
    online.) %
    Demonstration of off-axis vortices in the \gls{UFG} generated by firing a
    ``bullet'' along an axially symmetric trap in \movieref{a
      movie}~\cite{movie:science} from Ref.~\cite{Bulgac:2011b}.  Off-axis
    vortex rings attach to the walls, and turn into vortex lines that cross and
    reconnect.  \movieref{The movie}~\cite{movie:science} also demonstrates that
    vortex rings and lines move at similar velocities.}
\end{figure}

Gravity provides a systematic breaking of the axial symmetry, allowing for a
consistent conversion of the initial vortex rings into a horizontally oriented
vortex line, though the exact nature of this conversion seems to be quite
sensitive to geometry.  A similar sensitivity to geometry was noted in both
experiment~\cite{ZA-SSSK:2005lr} and theory~\cite{Bulgac:2011b} where a high
degree of axial symmetry was found to be critical in the formation and
stability of a vortex lattice.  As can be seen in \cite{Bulgac:2011b} (see the
movie~\cite{movie:science}), \cite{Scherpelz:2014}, and the simulations
presented here, vortex rings and vortices move with almost the same velocity.
This is to be expected since small segments of both vortex rings or vortex
lines move according to the well established Magnus relationship in response to
a similar buoyant force directed out from the trap (see~\cite{Bulgac:2013a} for
a discussion and references therein).  Both large rings and vortices have the
same structure and core depletion, and will therefore move with comparable
velocities and periods up to small corrections due to geometric (curvature)
effects.

Likewise, the expansion of a large vortex ring and a vortex segment will behave
similarly, so the subtle dependence of the final image on the imaging procedure
as discussed in~\cite{Bulgac:2013d} remains valid, explaining how the vortex
core of the inter-particle scale expands to appear as a much larger object. The
only potentially noticeable difference between the motion of vortex rings and
vortices is a possible asymmetry in the oscillations for vortex rings which
return as a small ring down the centre of the trap.  It turns out, however, that
the asymmetry in this motion is quite small except in the case of very large
amplitude oscillations~\cite{Bulgac:2013d, Pitaevskii:2013}.  As a result,
distinguishing between the two scenarios is best performed by imaging from
different directions or through tomography~\cite{Ku:2014}.

Here we extend the fermionic \gls{SLDA} simulations of~\cite{Bulgac:2013d} to
include an anisotropy $\delta$ and anharmonicity $C/C_0$ to model the dynamics
of an imprinted domain wall in a cloud of $\sim 560$ particles in the
\gls{UFG}.  We use the same formalism and initial state preparation detailed
in~\cite{Bulgac:2013d} on a $32^2\times 128$ lattice, but now include
gravitational effects~\eqref{eq:V_gravity} comparable to those describing the
trap in~\cite{Ku:2014}.

While our simulations only contain about a thousand particles, they provide a
more accurate representation of the initial experimental conditions than
available through any other technique.  In particular, the \gls{ETF} model -- a
bosonic \gls{GPE}-like \gls{DFT} -- used in~\cite{Bulgac:2013d, Reichl:2013,
  Wen:2013} lacks a mechanisms for the superfluid to relax.  Thus, while
suitable for studying the qualitative dynamics of vortex motion in large traps,
the \gls{ETF} is not suitable for the period shortly after the imprint where
the system exhibits significant relaxation (see the supplemental
material~\cite{EPAPS:apsSupp} for a comparison between the \gls{SLDA} and the
\gls{ETF}).  An \textit{ad hoc} relaxation was included in
Ref.~\cite{Scherpelz:2014}, but this model does not reproduce \gls{UFG}
equation of state, and lacks the quantitative validation of the \gls{SLDA}
where the functional is fully determined by fitting \gls{QMC} and experimental
results (see~\cite{Bulgac:2011} and references therein).  Despite the absence
of an explicit collision integral, the large number of degrees of freedom in
the \gls{SLDA} permit many mechanisms for superfluid relaxation including
various phonon processes, Cooper pair breaking, and Landau damping.

We start in Fig.~\ref{fig:strong_asym} with a moderate anisotropy $\delta
\approx 10\%$ which elongates the trap vertically as in the experiment.  As
demonstrated in~\cite{Bulgac:2013d}, the phase imprint rapidly seeds the
formation of a vortex ring, but as this ring evolves, it is attracted to the
surface and eventually hits the nearer boundaries on the side, converting to a
pair of vortex lines.  These lines then oscillate along the trap undergoing a
crossing and recombination process similar to that seen using \gls{GPE}
simulations~\cite{Piazza:2010} -- the mechanism responsible for quantum
turbulence -- changing orientation from horizontal to vertical and back.  The
addition of a $C/C_0 \approx 3\%$ anharmonicity along the $y$ direction, similar
to that induced by gravity in the experiment, breaks the mirror symmetry, and
one of the two vortices is gradually ejected from the system, leaving a single
vortex oriented horizontally along the shorter axis of the trap that oscillates
through the cloud, consistent with the observations~\cite{Ku:2014}.

\newcommand{\frameeq}[2]{
  \parbox[b]{\eqwidth}{\footnotesize
    \begin{align*}
      \delta&#1\\
      \frac{C}{C_0}&#2
    \end{align*}}}

\newlength{\eqwidth}
\setlength{\eqwidth}{\widthof{\footnotesize$\frac{C}{C_0}=0.15\%$}}

\begin{figure}[tb]
  \frameeq{\approx 10\%}{=0}%
  \href{http://www.youtube.com/watch?v=VxvOPsOs-6s&t=1}{\includegraphics{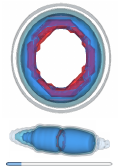}}%
  \href{http://www.youtube.com/watch?v=VxvOPsOs-6s&t=2}{\includegraphics{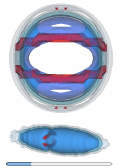}}%
  \href{http://www.youtube.com/watch?v=VxvOPsOs-6s&t=3}{\includegraphics{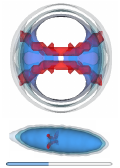}}%
  \href{http://www.youtube.com/watch?v=VxvOPsOs-6s&t=3}{\includegraphics{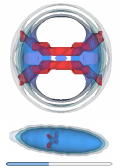}}%
  \href{http://www.youtube.com/watch?v=VxvOPsOs-6s&t=6}{\includegraphics{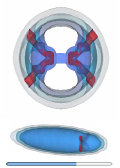}}%
  \href{http://www.youtube.com/watch?v=VxvOPsOs-6s&t=7}{\includegraphics{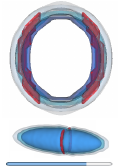}}%
  \\
  \frameeq{\approx 10\%}{\approx 3\%}%
  \href{http://www.youtube.com/watch?v=7qG03kmN-fs&t=1}{\includegraphics{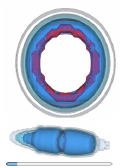}}%
  \href{http://www.youtube.com/watch?v=7qG03kmN-fs&t=2}{\includegraphics{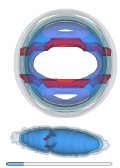}}%
  \href{http://www.youtube.com/watch?v=7qG03kmN-fs&t=3}{\includegraphics{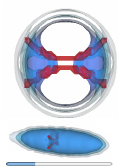}}%
  \href{http://www.youtube.com/watch?v=7qG03kmN-fs&t=3}{\includegraphics{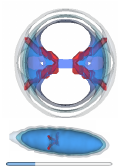}}%
  \href{http://www.youtube.com/watch?v=7qG03kmN-fs&t=6}{\includegraphics{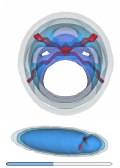}}%
  \href{http://www.youtube.com/watch?v=7qG03kmN-fs&t=7}{\includegraphics{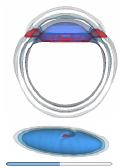}}
  \caption{\label{fig:strong_asym}(Color online, click on frames to view movie
    online.) Effects of moderate anisotropy $\delta = 1-\omega_y/\omega_x
    \approx 10\%$ (trap is taller than wide): the ring hits the narrow walls on
    the side, forming two parallel vortices.  Without any anharmonicity $C=0$
    (top), these undergo several recombinations, oscillating between a
    horizontal and vertical orientation.  The presence of an anharmonicity
    $C/C_0\approx 3\%$ breaks the symmetry, eventually expelling one vortex
    from the system, resulting in a long-lived vortex that oscillates back and
    forth.}
\end{figure}

\begin{figure}[tb]
  \frameeq{\approx 0.5\%}{=0}%
  \href{http://www.youtube.com/watch?v=gIfhB7QrGxU&t=1}{\includegraphics{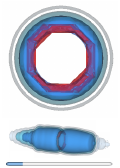}}%
  \href{http://www.youtube.com/watch?v=gIfhB7QrGxU&t=3}{\includegraphics{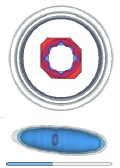}}%
  \href{http://www.youtube.com/watch?v=gIfhB7QrGxU&t=4}{\includegraphics{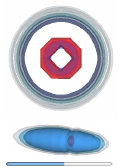}}%
  \href{http://www.youtube.com/watch?v=gIfhB7QrGxU&t=5}{\includegraphics{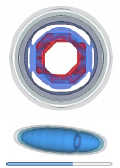}}%
  \href{http://www.youtube.com/watch?v=gIfhB7QrGxU&t=5}{\includegraphics{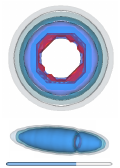}}%
  \href{http://www.youtube.com/watch?v=gIfhB7QrGxU&t=8}{\includegraphics{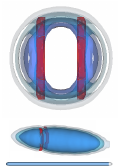}}%
  \\
  \frameeq{\approx 0.5\%}{\approx 0.15\%}%
  \href{http://www.youtube.com/watch?v=K74GUp1cDnM&t=1}{\includegraphics{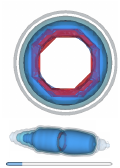}}%
  \href{http://www.youtube.com/watch?v=K74GUp1cDnM&t=3}{\includegraphics{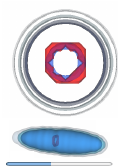}}%
  \href{http://www.youtube.com/watch?v=K74GUp1cDnM&t=4}{\includegraphics{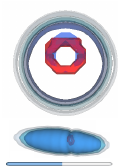}}%
  \href{http://www.youtube.com/watch?v=K74GUp1cDnM&t=5}{\includegraphics{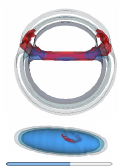}}%
  \href{http://www.youtube.com/watch?v=K74GUp1cDnM&t=5}{\includegraphics{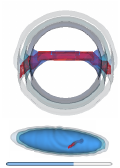}}%
  \href{http://www.youtube.com/watch?v=K74GUp1cDnM&t=8}{\includegraphics{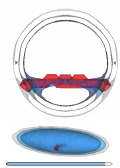}}%
  \caption{\label{fig:weak_asym}(Color online, click on frames to view movie
    online.) %
    Effects of weak anisotropy: the ring now lives long enough to return as a
    small ring.  Without an anharmonicity (top) several oscillations occur
    before the ring expands and hits the walls forming two parallel vortices.
    Adding a weak anharmonicity (bottom) causes the small ring to rotate.  Once
    twisted off axis, this ring rapidly converts to a single vortex as seen
    in~\cite{Scherpelz:2014}.  The rapid production of a single vortex from a
    twisted defect was also described in~\cite{Parker:2004} and provides a
    convenient way of constructing vortices.}
\end{figure}

In Fig.~\ref{fig:weak_asym} we compare similar systems, but now with
asymmetries an order of magnitude smaller.  In this case, the vortex ring
persists much longer before converting into vortex lines, long enough to return
as a small vortex down the centre of the trap.  This smaller vortex appears to
be very sensitive to even a tiny anharmonicity which causes it to tilt upwards
and collide with the upper wall of the trap, rapidly forming a single
horizontally aligned vortex. This rapid formation of a single vortex from a
tilted ring or imprint was also obtained in simulation~\cite{Scherpelz:2014}
and mentioned in~\cite{Parker:2004, Ku:2014}, and seems
like a more reliably mechanism for consistently forming a single horizontally
aligned vortex (see the movie tilted imprint in the supplemental
material~\cite{EPAPS:aps}).

\begin{figure}[tb]
  \frameeq{\approx -10\%}{\approx 3\%}%
  \href{http://www.youtube.com/watch?v=0pg8N91-gZ0&t=2}{\includegraphics{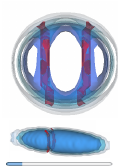}}%
  \href{http://www.youtube.com/watch?v=0pg8N91-gZ0&t=3}{\includegraphics{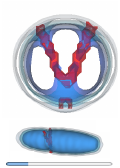}}%
  \href{http://www.youtube.com/watch?v=0pg8N91-gZ0&t=3}{\includegraphics{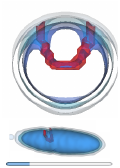}}%
  \href{http://www.youtube.com/watch?v=0pg8N91-gZ0&t=4}{\includegraphics{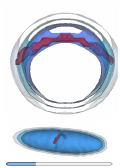}}%
  \\
  \frameeq{\approx 10\%}{\approx 3\%}%
  \href{http://www.youtube.com/watch?v=7xlG15TUNQs&t=1}{\includegraphics{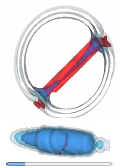}}%
  \href{http://www.youtube.com/watch?v=7xlG15TUNQs&t=2}{\includegraphics{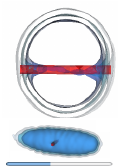}}%
  \href{http://www.youtube.com/watch?v=7xlG15TUNQs&t=4}{\includegraphics{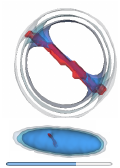}}%
  \href{http://www.youtube.com/watch?v=7xlG15TUNQs&t=6}{\includegraphics{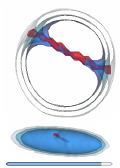}}%
  \caption{\label{fig:alignment}(Color online, click on frames to view movie
    online.) %
    The trap on the top is wider than it is tall -- the energetically favored
    configuration would be a vertical vortex, but the initial conditions lead
    to a slightly more energetic vortex aligned along the longer of the two
    axes. On the bottom, we imprint an oblique vortex line in a trap with the
    same asymmetry as the experiment.  As it oscillates long the trap, the
    vortex alignment oscillates about the favored horizontal position, but
    without significant damping, continues past this alignment to another
    oblique position.  The presence of significant damping -- such as through
    phonons at finite $T$ -- might allow this to relax to a horizontal
    alignment.  However this relaxation would likely be comparable to the time
    it takes the vortex to oscillate out of the system -- the true ``minimum''
    energy state has no vortex at all.}
\end{figure}

In particular, as demonstrated in the top of Fig.~\ref{fig:alignment},
even when given a negative anisotropy $\delta \approx -10\%$ so that
the trap is compressed vertically, a meta-stable horizontally aligned
vortex may still result.  It was suggested in~\cite{Ku:2014} that the
horizontal alignment results from the fact that this state has lower
energy, but evolution in these systems is conservative to a high
degree of accuracy, hence rapid progress toward a state with lower
energy or larger statistical weight is not guaranteed a priori.
Indeed, a lower energy state exists -- that with the vortex ejected
from the system -- and we expect the timescale for a vortex to relax into
a stable horizontal configuration to be comparable to or longer than
the timescale for a vortex to lose its energy and completely oscillate
out of the system.  To check, we imprinted a misaligned vortex in a
trap with both asymmetries $\delta\approx 10\%$ and $C/C_0\approx
3\%$.  As shown on the bottom of Fig.~\ref{fig:alignment}, although
this vortex does rotate toward the horizontal configuration, it
quickly rotates past the configuration to an approximate mirror
misalignment.

While here we have stressed the role of broken symmetries in the evolution
of vortex rings, it is still a challenge for experimentalists to perform similar
experiments in axially symmetric traps, similar to those used
in~\cite{ZA-SSSK:2005lr}, where vortex rings are likely to survive for long
periods of time.

Almost six decades ago, Feyman~\cite{Feynman:1955} envisioned vortex crossing
and recombination as responsible for quantum turbulence in cold superfluids
which lack dissipation at zero temperature.  While many study turbulent
phenomena~\cite{\citeturbulence}, we are aware of a few recent experiments that
have directly observed in experiment the vortex crossing and recombination
mechanism~\cite{Paoletti:2008, Bewley:2009, Guo:2014, Fonda:2014}. These
dynamics and interactions play a crucial role in many fermionic superfluids
with applications to conduced matter physics, neutron stars, cosmology,
particle physics (see e.g.~\cite{Volovik:2003}).  For example, to explain
pulsar glitches in neutron stars, one may need to quantitatively understand
energy loss during crossing and recombination as inputs to glitching models
(see e.g.~\cite{Bulgac:2013a}).  The \gls{UFG} provides an almost ideal
laboratory study these phenomena and benchmark the \gls{SLDA}.  Using multiple
tilted imprints, for example, one can control the generation and arrangement of
multiple vortices in order to study collisions, reconnection, and
interactions. The \gls{UFG} and \gls{SLDA} thus provides a new microscopic
framework to study aspects of quantum turbulence in a strongly
interacting system, complementing weakly-interacting dilute Bose
gases~\cite{Weiler:2008, Henn:2009} modelled with the \gls{GPE} as the only
microscopic frameworks presently available for studying superfluid dynamics.

\begin{figure}[tb]
  \href{http://www.youtube.com/watch?v=nCGtq8k6SAQ&t=0}{\includegraphics{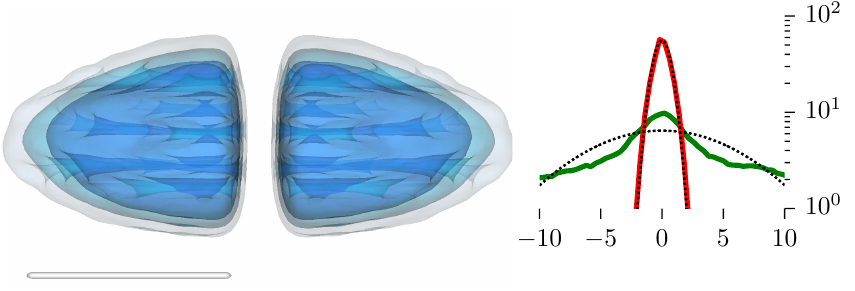}}
  \\[-5pt]
  \href{http://www.youtube.com/watch?v=nCGtq8k6SAQ&t=1}{\includegraphics{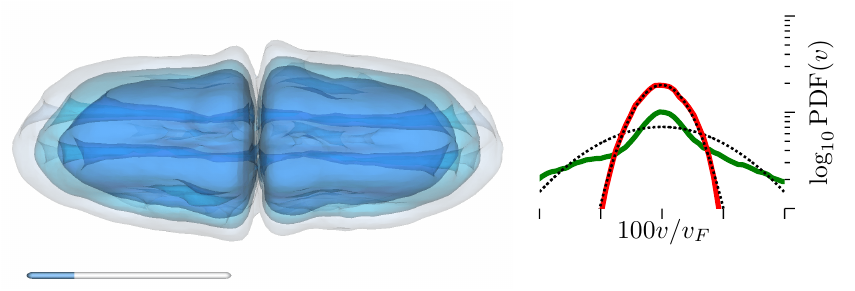}}
  \\[-5pt]
  \href{http://www.youtube.com/watch?v=nCGtq8k6SAQ&t=4}{\includegraphics{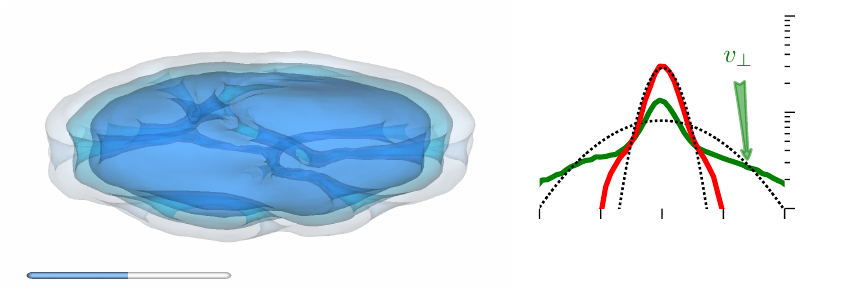}}
  \\[-5pt]
  \href{http://www.youtube.com/watch?v=nCGtq8k6SAQ&t=6}{\includegraphics{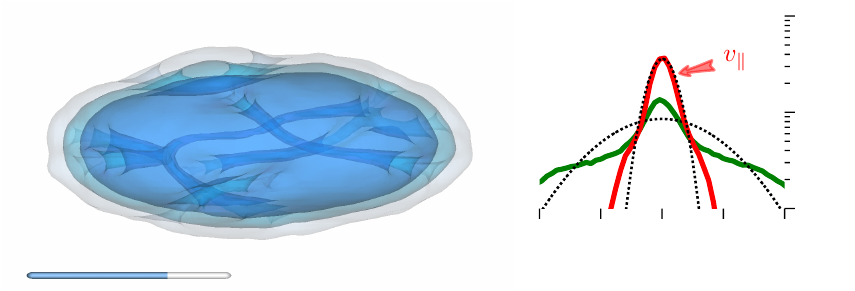}}
  \\[-5pt]
  \href{http://www.youtube.com/watch?v=nCGtq8k6SAQ&t=7}{\includegraphics{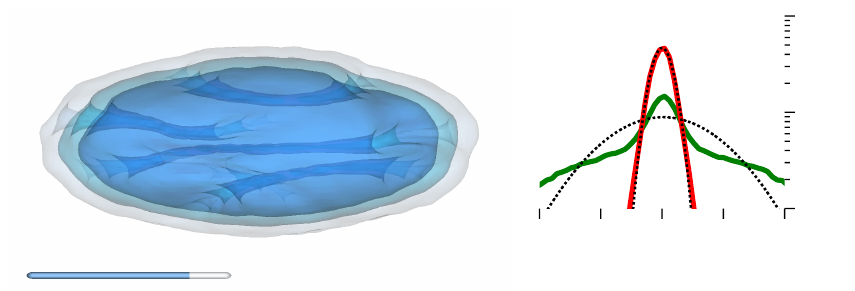}}
  \caption{\label{fig:qt_gen}(Color online, click on frames to view movie
    online.) Generation of quantum turbulence by phase imprint of the vortex
    lattice. In the left column consecutive frames show: a) vortex lattice with
    knife edge dividing cloud, b) just after phase imprint removal of the
    knife, c-e) decay of turbulent motion.  In the right column we show the
    corresponding \glspl{PDF} for longitudinal $v_{||}$ and transverse
    $v_\perp$ components of collective velocity. Dotted lines show the gaussian
    best fit to the data.}
\end{figure}

The phase imprint technique can be also utilized to create turbulent states
with many tangled vortices.  Here we demonstrate one approach, adding a phase
imprint~\cite{Yefsah:2013, Ku:2014} to a lattice of vortices which can be
created experimentally by stirring using laser beams~\cite{ZA-SSSK:2005lr}.  In
Fig.~\ref{fig:qt_gen} we show consecutive frames of turbulent motion exhibiting
crossings and recombinations of quantized vortices in an elongated harmonic
trap. The simulation was done in a $48^2 \times 128$ box comprising $1410$
fermions (see supplemental material~\cite{EPAPS:aps} for a movie).  We also
show the corresponding \gls{PDF} of the velocities for longitudinal
$v_{\parallel}$ and transverse $v_{\perp}$ components of the velocity (with
respect to long axis).

We start with the ground state of a cloud cut in half with a knife-edge
potential. We then stir the system with two circulating laser beams parallel to
the long axis of the trap.  Once a vortex lattice is generated, we imprint a
$\pi$ phase shift between the halves.  Just before removing the edge knife, we
introduce a slight tilt to speed the formation of a vortex tangle.  After the
knife-edge is removed, the vortex lines twist, cross, and reconnect.  From the
velocity \glspl{PDF} one sees a clear departure from gaussian behaviour as the
tangle evolves -- a hallmark of quantum turbulence.  Eventually the system
relaxes to a vortex lattice and equilibrates in $v_\parallel$ .  Somewhat
similar velocity \glspl{PDF} are seen in theoretical studies of dilute Bose
gases~\cite{White:2010} and in phenomenological filament model of the
crossing-recombination vortex line dynamics~\cite{Adachi:2011}.

In conclusion, have shown the crucial role played by the trap geometry in the
formation of a vortex line after a phase imprint. In particular we identified a
few possible scenarios for the short term evolution of the phase imprint in the
experiments~\cite{Yefsah:2013, Ku:2014}, showing that the details are highly
sensitive to geometric factors. To precisely characterize the behavior realized
in the experiments~\cite{Yefsah:2013, Ku:2014}, the experiment will likely need
to be simulated with precise values of the trapping asymmetries known, and with
realistic particle numbers which are currently are beyond the capabilities of
the most advanced implementations of the \gls{SLDA} approach.  Satisfactory
agreement with the latest MIT experiments serves as the next step in validating
the time-dependent \gls{SLDA}, demonstrating that it is capable of
qualitatively describing the complex dynamics of strongly interacting fermionic
systems.  We have demonstrated that recombination is likely present in the
early stages of the experiments~\cite{Yefsah:2013, Ku:2014} (see
Figs.~\ref{fig:strong_asym} and \ref{fig:weak_asym}) and can be selected for by
reducing the anharmonicity of the trap. We have also presented that the phase
imprint technique can be utilized to generate quantum turbulent state.
Therefore, using improved imaging techniques~\cite{Ku:2014} coupled with
carefully designed initial conditions, cold atom experiments have a great
opportunity to directly probe and quantify the dynamics and interactions
vortices and the potential to significantly advance our understanding of
quantum turbulence. In this regard, the unitary Fermi gas is of particular
interest as the results will have almost direct impact on superfluid dynamics
and turbulent phenomena in strongly interacting Fermi superfluids, in
particular neutron star phenomenology.

We thank L.~Chuek, M.~Ku, and M.~Zwierlein for describing details of their
experiment.  We acknowledge support under U.S. Department of Energy (DoE) Grant
No. DE-FG02-97ER41014 and the Polish National Science Center (NCN) grant under
decision No.\@ DEC-2013/08/A/ST3/00708.  G.W.\@ acknowledges the Center for
Advanced Studies at Warsaw University of Technology for the support under
Contract No. 58/2013 (international research scholarships financed by the
European Union from the European Social Funds, CAS/32/POKL).  MMF acknowledges
support from the Institute for Nuclear Theory during the program
\textit{Universality in Few-Body Systems: Theoretical Challenges and New
  Directions, INT-14-1}.  Some of the calculations reported here have been
performed at the University of Washington Hyak cluster funded by the NSF MRI
Grant No.\@ PHY-0922770. This research also used resources of the National
Center for Computational Sciences at Oak Ridge National
Laboratory~\cite{titan}, which is supported by the Office of Science of the DoE
under Contract DE-AC05-00OR22725.

%

\clearpage
\newpage

\section*{Supplemental Material}\noindent
In this supplement we provide detailed information about the simulations, list
of attached movies as well as comparison of the \gls{SLDA} simulations with
the corresponding \gls{ETF} simulation.

\subsection{Simulation parameters}\noindent
Simulations are done in a box of size $n_x\times n_y \times n_z$, where
$n_x=n_y=32$ and $n_z=128$.  The simulations contain $N\approx560$ particles in
the following trapping potential:
\begin{equation*}
  V(x,y,z)=\dfrac{m}{2}\omega_x^2x^2+\dfrac{m}{2}\omega_y^2y^2
  + \dfrac{m}{2}\omega_z^2z^2 + Cy^3,
\end{equation*}
where aspect ratio of the trap $\omega_\bot / \omega_z = 4$
($\omega_\bot=\sqrt{\omega_x \omega_y}$) is fixed.  The remaining trap
parameters are characterized by an anisotropy: $\delta = 1-\omega_y/\omega_x$,
and an anharmonicity: $C/C_0$, where $C_0 = m\omega_y^2/2R$ and $R$ is
Thomas-Fermi radius. The simulation requires to evolve $57,849$ two component wave-functions in real time.
In order to integrate equations of motion we
use a symplectic split-operator method that respects time-reversal symmetry
launched on hundreds of GPUs on the Titan supercomputer.

\subsection{Movies -- Fermionic Simulations}\noindent
Movies show paring field profiles (blue surfaces) and location of a vortex (red
line) identified as points around which phase of the paring field rotates by
$2\pi$.

\begin{description}
\item[\texttt{A.mp4}] [\utube{\movieA}]\\
  Parameters: $\delta=9\%$, $C/C_0 = 0\%$.
\item[\texttt{B.mp4}] [\utube{\movieB}]\\
  Parameters: $\delta=0.5\%$, $C/C_0 = 0\%$.
\item[\texttt{C.mp4}] [\utube{\movieC}]\\
  Parameters: $\delta=9\%$, $C/C_0 = 3\%$.
\item[\texttt{D.mp4}] [\utube{\movieD}]\\
  Parameters: $\delta=0.5\%$, $C/C_0 = 0.15\%$.
\item[\texttt{E.mp4}] [\utube{\movieE}]\\
  Parameters: $\delta=-10\%$, $C/C_0 = 3\%$.
\item[\texttt{F.mp4}] [\utube{\movieF}]\\
  Parameters: $\delta=0\%$, $C/C_0 = 0\%$.\\
  \textit{Comment}: The edge knife is tilted by angle $\pi/20\,\rm{rad}$.
\item[\texttt{G.mp4}] [\utube{\movieG}]\\
  Parameters: $\delta=0\%$, $C/C_0 = 0\%$.\\
  \textit{Comment}: Simulation with two edge knifes tilted by angle
  $\pi/20\,\rm{rad}$ with different orientation.
\item[\texttt{H.mp4}] [\utube{\movieH}]\\
  Parameters: $\delta=9\%$, $C/C_0 = 3\%$.\\
  \textit{Comment}: The edge knife is tilted by angle $\pi/20\,\rm{rad}$ and
  rotated in a such way to generate oblique vortex line.
\end{description}

\subsection{Movies -- Bosonic Simulations (extended Thomas-Fermi model)}
\noindent
Movies show order parameter $2|\Psi|$ profiles (blue surfaces) and location of a vortex (red line) identified as
points around which phase of the order parameter rotates by $2\pi$. Results are obtained using the same lattice parameters as in case of fermionic simulations.

\begin{description}
\item[\texttt{A.ETF.mp4}] [\utube{\movieETFA}]\\
  Parameters: $\delta=9\%$, $C/C_0 = 0\%$.
\item[\texttt{B.ETF.mp4}] [\utube{\movieETFB}]\\
  Parameters: $\delta=0.5\%$, $C/C_0 = 0\%$.
\item[\texttt{C.ETF.mp4}] [\utube{\movieETFC}]\\
  Parameters: $\delta=9\%$, $C/C_0 = 3\%$.
\item[\texttt{D.ETF.mp4}] [\utube{\movieETFD}]\\
  Parameters: $\delta=0.5\%$, $C/C_0 = 0.15\%$.
\item[\texttt{E.ETF.mp4}] [\utube{\movieETFE}]\\
  Parameters: $\delta=-10\%$, $C/C_0 = 3\%$.
\end{description}

\subsection{Simulation of Quantum Turbulence}\noindent
The simulation of quantum turbulence was done with lattice $48 \times 48 \times
128$, and particle number $N=1410$.  The confining potential is axially
symmetric ($\delta=0\%$) with aspect ratio $\omega_\bot / \omega_z =
2.67$. There is no anharmonicity term ($C/C_0 = 0\%$).
Number of two component wave-functions to be evolved in real time is $131,629$
and to integrate equations of motion we used 2048 Titan's GPUs.
As the stirrers we used
repulsive Gaussian potentials of width $1.25$ of lattice spacing and amplitude
$0.75\varepsilon_F$. We provide two movies:
\begin{description}
\item[\texttt{QT.mp4}] [\utube{\movieQT}]\\
  Dynamics of the cloud showing paring field profiles (blue surfaces).
\end{description}

\subsection{Comparison of Fermionic and Bosonic Simulations}\noindent
On the next page we compare the \gls{SLDA} simulations presented in the main
body of the paper (top of each figure) with the corresponding \gls{ETF}
simulation with the same initial conditions and preparation.
\clearpage

\begin{figure}[H]
  \frameeq{\approx 10\%}{=0}%
  \href{http://www.youtube.com/watch?v=VxvOPsOs-6s&t=1}{\includegraphics{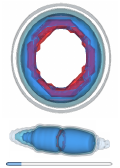}}%
  \href{http://www.youtube.com/watch?v=VxvOPsOs-6s&t=2}{\includegraphics{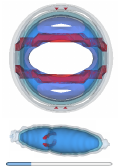}}%
  \href{http://www.youtube.com/watch?v=VxvOPsOs-6s&t=3}{\includegraphics{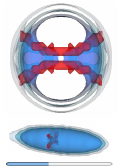}}%
  \href{http://www.youtube.com/watch?v=VxvOPsOs-6s&t=3}{\includegraphics{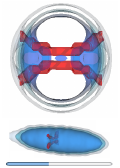}}%
  \href{http://www.youtube.com/watch?v=VxvOPsOs-6s&t=6}{\includegraphics{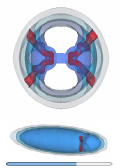}}%
  \href{http://www.youtube.com/watch?v=VxvOPsOs-6s&t=7}{\includegraphics{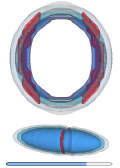}}%
  \\[1em]
  \frameeq{\approx 10\%}{=0}%
  \href{http://www.youtube.com/watch?v=uDuOOVOO5D4&t=1}{\includegraphics{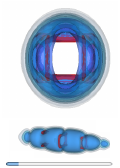}}%
  \href{http://www.youtube.com/watch?v=uDuOOVOO5D4&t=2}{\includegraphics{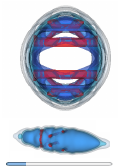}}%
  \href{http://www.youtube.com/watch?v=uDuOOVOO5D4&t=3}{\includegraphics{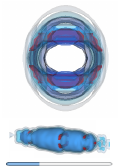}}%
  \href{http://www.youtube.com/watch?v=uDuOOVOO5D4&t=3}{\includegraphics{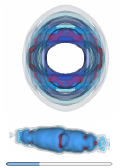}}%
  \href{http://www.youtube.com/watch?v=uDuOOVOO5D4&t=6}{\includegraphics{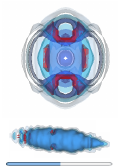}}%
  \href{http://www.youtube.com/watch?v=uDuOOVOO5D4&t=7}{\includegraphics{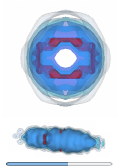}}
  \caption{\label{fig:ETFstrong_a}%
    (Color online, click on frames to view movie online.) %
    \Gls{ETF} analogue of the \gls{SLDA} simulation in the top of
    Fig.~\ref{fig:strong_asym}}
\end{figure}

\begin{figure}[H]
  \frameeq{\approx 10\%}{\approx 3\%}%
  \href{http://www.youtube.com/watch?v=7qG03kmN-fs&t=1}{\includegraphics{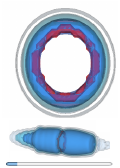}}%
  \href{http://www.youtube.com/watch?v=7qG03kmN-fs&t=2}{\includegraphics{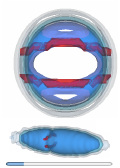}}%
  \href{http://www.youtube.com/watch?v=7qG03kmN-fs&t=3}{\includegraphics{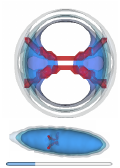}}%
  \href{http://www.youtube.com/watch?v=7qG03kmN-fs&t=3}{\includegraphics{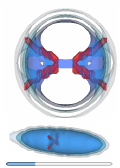}}%
  \href{http://www.youtube.com/watch?v=7qG03kmN-fs&t=6}{\includegraphics{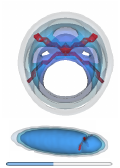}}%
  \href{http://www.youtube.com/watch?v=7qG03kmN-fs&t=7}{\includegraphics{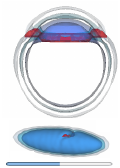}}%
  \\[1em]
  \frameeq{\approx 10\%}{\approx 3\%}%
  \href{http://www.youtube.com/watch?v=L5ONtMQDfgE&t=1}{\includegraphics{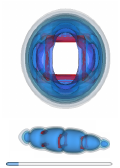}}%
  \href{http://www.youtube.com/watch?v=L5ONtMQDfgE&t=2}{\includegraphics{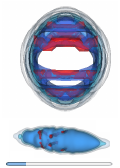}}%
  \href{http://www.youtube.com/watch?v=L5ONtMQDfgE&t=3}{\includegraphics{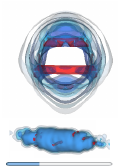}}%
  \href{http://www.youtube.com/watch?v=L5ONtMQDfgE&t=3}{\includegraphics{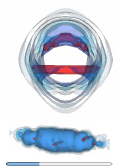}}%
  \href{http://www.youtube.com/watch?v=L5ONtMQDfgE&t=6}{\includegraphics{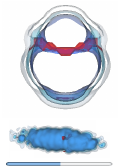}}%
  \href{http://www.youtube.com/watch?v=L5ONtMQDfgE&t=7}{\includegraphics{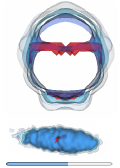}}
  \caption{\label{fig:ETFstrong_b}%
    (Color online, click on frames to view movie online.) %
    \Gls{ETF} analogue of the \gls{SLDA} simulation in the bottom of
    Fig.~\ref{fig:strong_asym}}
\end{figure}

\begin{figure}[H]
  \frameeq{\approx 0.5\%}{=0}%
  \href{http://www.youtube.com/watch?v=gIfhB7QrGxU&t=1}{\includegraphics{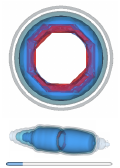}}%
  \href{http://www.youtube.com/watch?v=gIfhB7QrGxU&t=3}{\includegraphics{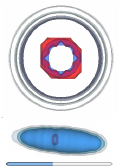}}%
  \href{http://www.youtube.com/watch?v=gIfhB7QrGxU&t=4}{\includegraphics{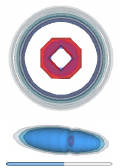}}%
  \href{http://www.youtube.com/watch?v=gIfhB7QrGxU&t=5}{\includegraphics{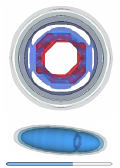}}%
  \href{http://www.youtube.com/watch?v=gIfhB7QrGxU&t=5}{\includegraphics{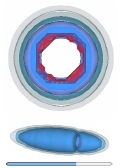}}%
  \href{http://www.youtube.com/watch?v=gIfhB7QrGxU&t=8}{\includegraphics{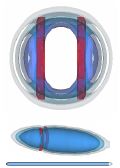}}%
  \\[1em]
  \frameeq{\approx 0.5\%}{=0}%
  \href{http://www.youtube.com/watch?v=VUrlvJglqkI&t=1}{\includegraphics{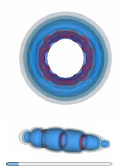}}%
  \href{http://www.youtube.com/watch?v=VUrlvJglqkI&t=3}{\includegraphics{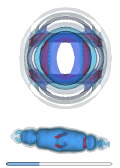}}%
  \href{http://www.youtube.com/watch?v=VUrlvJglqkI&t=4}{\includegraphics{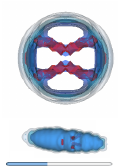}}%
  \href{http://www.youtube.com/watch?v=VUrlvJglqkI&t=5}{\includegraphics{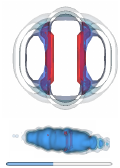}}%
  \href{http://www.youtube.com/watch?v=VUrlvJglqkI&t=5}{\includegraphics{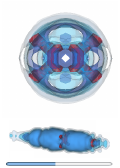}}%
  \href{http://www.youtube.com/watch?v=VUrlvJglqkI&t=8}{\includegraphics{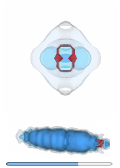}}
  \caption{\label{fig:ETFweak_a}%
    (Color online, click on frames to view movie online.) %
    \Gls{ETF} analogue of the \gls{SLDA} simulation in the top of
    Fig.~\ref{fig:weak_asym}}%
\end{figure}

\begin{figure}[H]
  \frameeq{\approx 0.5\%}{\approx 0.15\%}%
  \href{http://www.youtube.com/watch?v=K74GUp1cDnM&t=1}{\includegraphics{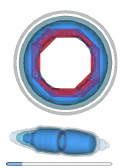}}%
  \href{http://www.youtube.com/watch?v=K74GUp1cDnM&t=3}{\includegraphics{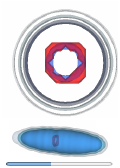}}%
  \href{http://www.youtube.com/watch?v=K74GUp1cDnM&t=4}{\includegraphics{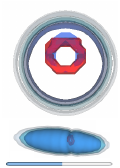}}%
  \href{http://www.youtube.com/watch?v=K74GUp1cDnM&t=5}{\includegraphics{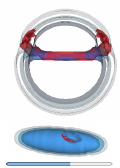}}%
  \href{http://www.youtube.com/watch?v=K74GUp1cDnM&t=5}{\includegraphics{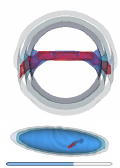}}%
  \href{http://www.youtube.com/watch?v=K74GUp1cDnM&t=8}{\includegraphics{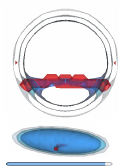}}%
  \\[1em]
  \frameeq{\approx 0.5\%}{\approx 0.15\%}%
  \href{http://www.youtube.com/watch?v=HOIB-R1TwJU&t=1}{\includegraphics{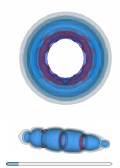}}%
  \href{http://www.youtube.com/watch?v=HOIB-R1TwJU&t=3}{\includegraphics{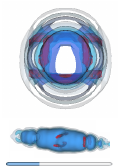}}%
  \href{http://www.youtube.com/watch?v=HOIB-R1TwJU&t=4}{\includegraphics{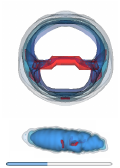}}%
  \href{http://www.youtube.com/watch?v=HOIB-R1TwJU&t=5}{\includegraphics{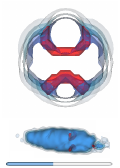}}%
  \href{http://www.youtube.com/watch?v=HOIB-R1TwJU&t=5}{\includegraphics{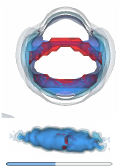}}%
  \href{http://www.youtube.com/watch?v=HOIB-R1TwJU&t=8}{\includegraphics{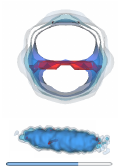}}
  \caption{\label{fig:ETFweak_b}%
    (Color online, click on frames to view movie online.) %
    \Gls{ETF} analogue of the \gls{SLDA} simulation in the bottom of
    Fig.~\ref{fig:weak_asym}}%
\end{figure}

\begin{figure}[H]
  \frameeq{\approx -10\%}{\approx 3\%}%
  \href{http://www.youtube.com/watch?v=0pg8N91-gZ0&t=2}{\includegraphics{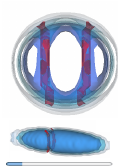}}%
  \href{http://www.youtube.com/watch?v=0pg8N91-gZ0&t=3}{\includegraphics{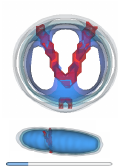}}%
  \href{http://www.youtube.com/watch?v=0pg8N91-gZ0&t=3}{\includegraphics{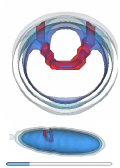}}%
  \href{http://www.youtube.com/watch?v=0pg8N91-gZ0&t=4}{\includegraphics{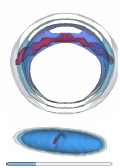}}%
  \\[1em]
  \frameeq{\approx -10\%}{\approx 3\%}%
  \href{http://www.youtube.com/watch?v=EXLOlBbsKS8&t=2}{\includegraphics{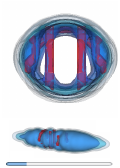}}%
  \href{http://www.youtube.com/watch?v=EXLOlBbsKS8&t=3}{\includegraphics{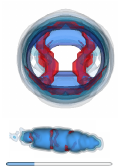}}%
  \href{http://www.youtube.com/watch?v=EXLOlBbsKS8&t=3}{\includegraphics{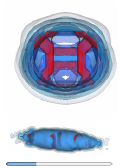}}%
  \href{http://www.youtube.com/watch?v=EXLOlBbsKS8&t=4}{\includegraphics{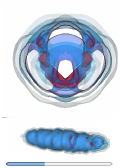}}
  \caption{\label{fig:ETFalign}%
    (Color online, click on frames to view movie online.) %
    \Gls{ETF} analogue of the \gls{SLDA} simulation in the top of
    Fig.~\ref{fig:alignment}}
\end{figure}
\end{document}